# The 1m$^3$ Semidigital Hadronic Prototype

M.C Fouz[1] for the CALICE Collaboration

1 – Centro de Investigaciones Energéticas, Medioambientales y Tecnológicas – Departamento de Investigación Básica
Avda. Complutense 40, 28040 Madrid – Spain

A high granularity hadronic 1 m$^3$ calorimeter prototype with semi-digital readout has been designed and built. This calorimeter has been made using stainless steel as absorber and Glass Resistive Plates Chambers (GRPC) as active medium, and read out through 1x1 cm$^2$ pads. This prototype aims to demonstrate that this technology fulfills the physics requirements for future linear collider experiments, and also to test the feasibility of building a realistic detector, taking into account design aspects as for instance a fully embedded front-end electronics based on power pulsing system, a compact and self-supporting mechanical structure, one-side services…

## 1 Introduction

Many interesting physics processes in a future lepton collider will involve multi-jet final states, often accompanied by charged leptons or missing energy. In order to exploit the rich physics potential of this collider, an improvement of the jet resolution of the order of a factor two compared with what has been achieved in previous experiments will be needed. In order to achieve this, one should think of new approaches instead of the ones from traditional calorimetry, and one of the most promising techniques is based on Particle Flow Algorithms (PFA) [1]. This concept, in contrast with a purely calorimetric measurement, requires the reconstruction of all the particles in the event, both charged and neutral ones. Particles are tracked in different sub-detectors and their energy is estimated from the most precise measurement (charged particles are measured by the tracker, photons by the electromagnetic calorimeter, and the neutral hadrons by using the information of the hadronic calorimeter and the electromagnetic calorimeter). The jet resolution obtained is a combination of the detector information and the reconstruction software.

This represents a new approach to calorimeters; they must not only measure the particle energy but, in addition, they should have a strong tracking capability to separate the contributions from the individual particles belonging to a jet. High granularity, both in longitudinal and transverse directions, becomes even more important than the energy resolution in order to make a correct assignment of calorimeter hits to the charged particles, and in order to have an efficient discrimination of nearby showers.

To improve the tracking, those calorimeters should be placed within a magnetic field. This adds constraints on materials and on the available space.

Due to the huge number of electronics channels, the read out electronics must be embedded in the detector and, to reduce the power consumption, it should operate in power pulsing mode matching the LC beam time cycles.

## 2 The Semi-Digital Hadronic Calorimeter concept

The CALICE collaboration [2] is developing highly segmented electromagnetic and hadronic calorimeters using different technologies. One of them is a semi-digital hadron calorimeter (SDHCAL) using stainless steel as absorber and a gas detector, read by pads of 1x1 $cm^2$, as active medium. Two options are considered. Glass Resistive Plate Chambers (GRPCs) and MICROMEGAS. Currently the GRPC is the baseline.

The SDHCAL incorporates a new concept: instead of recording the deposited energy on the calorimeter, it will register in how many pads, and in which ones, energy bigger than a certain threshold has been deposited. A semi-digital readout, with two different thresholds, is used to improve the linearity and resolution at high energies with respect to a purely digital option, due to a mitigation of the saturation effects. The semi-digital approach reduces the complexity and costs of the electronics.

To validate the SDHCAL concept a 1 $m^3$ prototype has been built. The prototype is intended to come as close as possible to a hadron calorimeter of the future ILC experiments This prototype aims to demonstrate that this technology fulfills the physics requirements but also the feasibility of building a realistic detector taking into account electronics and mechanical design aspects as a fully embedded front-end electronics based on power pulsing system, a compact and robust self-supporting mechanical structure, one-side services… The construction of this prototype allows gain experience with the procedures and possible problems in view of developing the final design. The next sections describe the design and construction of this prototype and show the first events collected during the start-up of commissioning with beam particles at CERN.

## 3 The Glass Resistive Parallel Plates

The gas detectors used in this prototype are Glass Resistive Parallel Plates (GRPC). The GRPC is a simple and robust gaseous detector consisting of two planes of highly resistive glass plates separated a few millimeters by thin spacers and filled with gas (a mixture of TFE/i-C4H10 (or CO2) /SF6 (93-94.5) / 5 / (2-0.5)). Figure 1 shows the cross section of the GRPCs used in this prototype and the associated electronics. The thinner glass (0.7mm) is used to build the anode while the thicker one (1.1 mm) forms the cathode. The anode thickness has been reduced with the purpose of reducing the multiplicity of the signal, i.e. the number of readout pads fired by a particle. The plates are kept apart 1.2 mm by ceramic balls (diameter 1.2 mm) and cylindrical buttons (diameter 4 mm). The advantage of these spacers with respect to using fishing lines is the reduction of the dead zones (0.1% versus few percent) and noise.

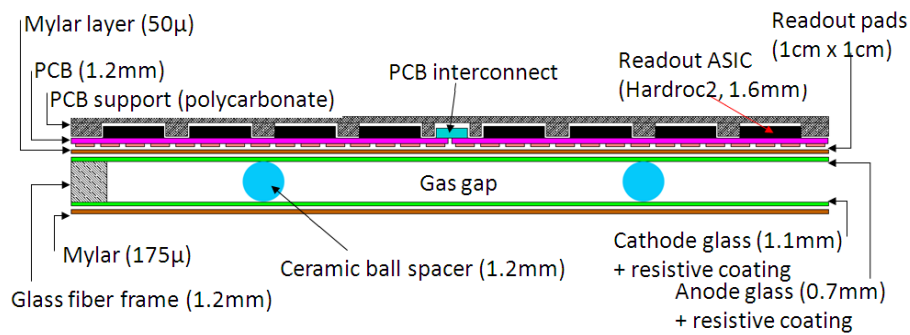

Figure 1: Cross section of a GRPC and its electronics readout.

The gas volume is closed by a glass fiber frame of 1.2 mm thick and 3 mm wide glued on both glass plates. The outer sides of the glass plates are covered by a resistive coating which is used to apply the high voltage. A special effort has been done to evaluate the best coating for the electrodes, taking into account the pad multiplicity, the homogeneity, and painting procedures. It was found that using the silk screen method it is possible to obtain a very uniform surface quality with stable resistivity in the range 0.5-2 MΩ/Square

A 50 microns thick Mylar layer separates the anode from the $1x1$ cm$^2$ copper pads of the electronic board that will be described in the next section. The pads pick up the signal induced by the charge of the avalanche electrons caused by the ionization in the gas.

## 4  The read out electronics

A chip called HARDROC (HARdronic Detector ReadOut Chip) [3] is used for the signal read out. Each chip contains 64 channels, it has three adjustable thresholds and a digital memory to store up to 127 events. The gain of each channel can be adjusted separately by a factor between 0 and 4 with a 6-bit precision. Each channel of the ASIC has a test capacitor of 2±0.02 pF which can be used to calibrate its response. It is very useful to have a uniform response for all channels. The HARDROC is also equipped with a power pulsing system to reduce the power dissipation by a factor 100 in the case of the proposed ILC duty cycle (their consumption is lower than 10 μW/channel).

Due to the high granularity of the SDHCAL a real calorimeter made with this technique will have more than 50 millions of read out channels. A single detector plate of 1 m$^2$ needs about ten thousand channels. The very front-end part of the readout electronics is integrated in the detector itself. A printed circuit board hosts the $1x1$ cm$^2$ copper pads for the GRPC read out and the HARDROC chips in the opposite face. The pads are connected to the HARDROC channels through the board structure. This board also provides the connection between adjacent chips and links the first to the readout system.

Due to some difficulties on producing and handling a 1m$^2$ board, it was decided to use 6 smaller boards. Each of these boards (called ASU=Active Sensor Unit) hosts 24 readout ASIC chips. The HARDROC are controlled by specifically Detector InterFace (DIF) cards. Every two PCBs are connected to each other and connected to one DIF. The DIF powers the ASUs, distributes the DAQ commands to the chips and transmits the collected data. Figure 2 shows a picture of the electronics used to read out a square meter GRPC made of 6 ASU hosting the 6x24 HARDROC connected to the 3 DIFs.

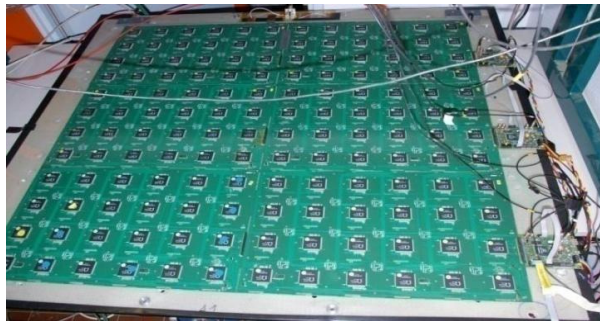

Figure 2: Picture of the electronics of 1m$^2$ GRPC. It can be seen the 6 ASUS connected to 3 DIF.

During the assembly of the SDHCAL prototype all the different electronic components have been verified before and after have been assembled. About ten thousand HARDROC chips have been tested in a special test bench using a Labview based application. The system was automated using a robot arm to pick-up the chip and place it in the test area. Different tests were performed to check the DC levels and power consumption, the slow control loading, linearity, trigger efficiency and memory.

# 5 The 1m3 SDHCAL Prototype

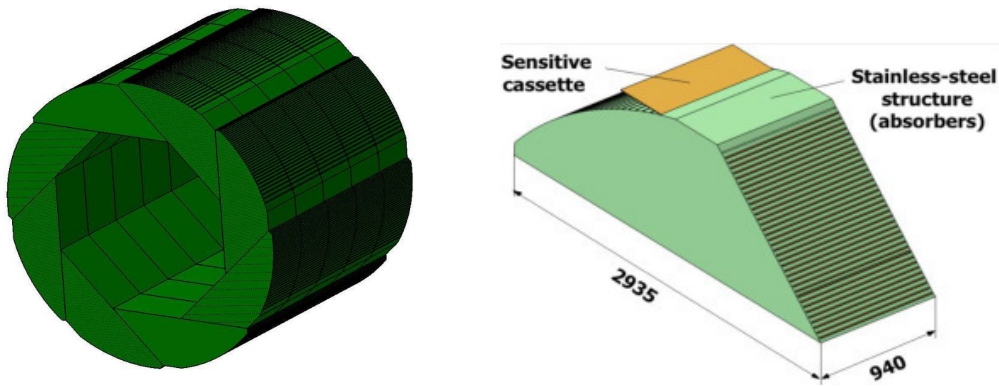

Figure 3: Design layout of the ILD Barrel HCAL (left) and one single module (right).

## 5.1 The Design

Figure 3 (left) shows the design layout for the barrel part of the HCAL proposed in the framework of the International Large Detector (ILD) [4]. The design has been optimized to reduce cracks. The barrel consists of 5 wheels each made of 8 identical modules. Each module is made of 48 stainless steel absorber plates interleaved with detector cassettes of different sizes as showed in Figure 3 (right).

Since the geometry of these modules is not appropriate for studying the performance of this calorimeter in test beam, a simpler geometry has been adopted for the prototype: a cubic design with all plates and detectors having the same dimensions of ~1x1 m$^2$. This allows a much better coverage of the hadronic shower profile. Figure 4 shows the prototype design. It consists on a mechanical structure, made of the absorber plates, that hosts the detector cassettes. The design allows an easy insertion and further extraction of the cassettes. The mechanical structure is made of 51 stainless steel plates assembled together using lateral spacers fixed to the absorbers through staggered bolts as can be seen in Figure 5. The dead spaces have been minimized as much as possible taking into account the mechanical tolerances (lateral dimensions and planarity) of absorbers and cassettes, to ensure a safe insertion/extraction of the cassette.

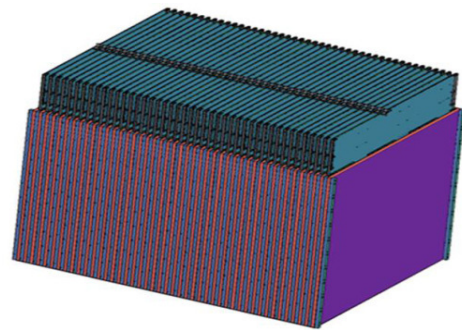

Figure 4: 1m$^3$ SDHCAL prototype design.

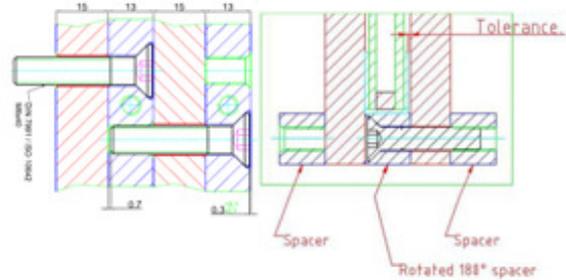

Figure 5: Details of lateral (left) and top (right) corners of the design. It shows some assembly details including the plates, spacers and bolts.

The plate dimensions are $1011 \times 1054 \times 15$ mm$^3$. The thickness tolerance is 0.05 mm and a surface planarity below ~500 microns was required. The spacers are 13 mm thick with 0.05 mm accuracy. The excellent accuracy of plate planarity and spacer thickness allowed reducing the tolerances needed for the safe insertion of the detectors. This is important to minimize the dead spaces and reduce the longitudinal size view of a future real detector.

The thickness and flatness of the plates used for the prototype has been verified using a laser interferometer system in order to certify they were inside the tolerances. The Figure 6 shows, as an example, the planarity distribution for one face of one of the plates. For this particular plate the maximum deviation from planarity is lower than 150 microns. For most plates the maximum does not exceed the required 500 microns. Figure 7 shows the maximum deviations from planarity for the first 44 measured plates.

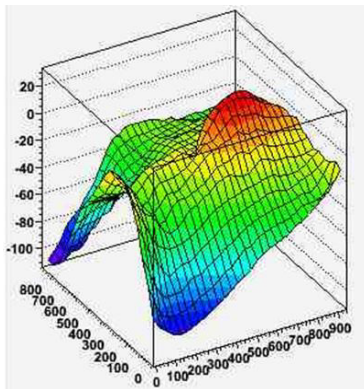

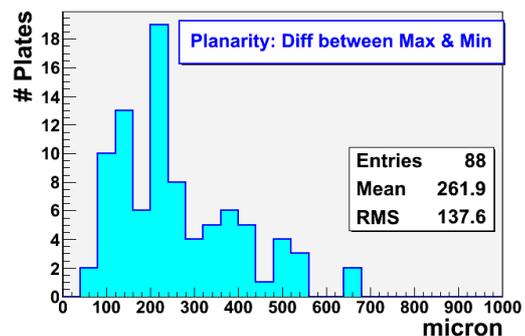

Figure 6: Planarity distribution for one side of one of the plates. Vertical axis units are microns, horizontal axis units are mm.

Figure 7: Maximum planarity differences for the first 44 plates (measurements from both faces).

## 5.2 Assembly of the mechanical structure

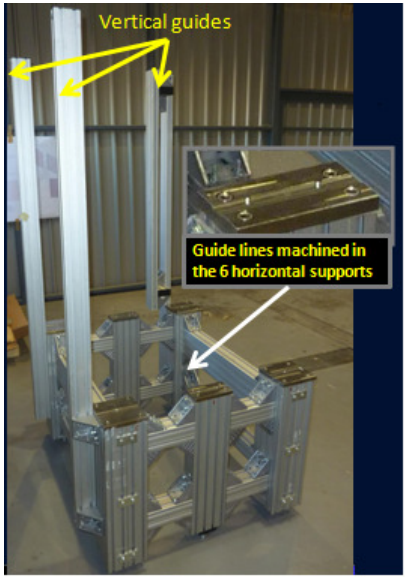

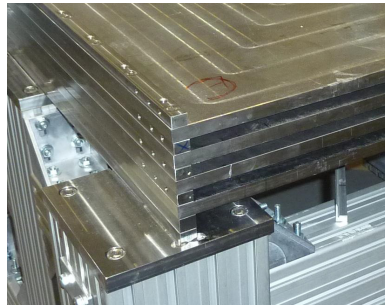

For the assembly of the mechanical structure a special table has been designed and built at CIEMAT (see Figure 8). This table must support a weight of about 6 tons. The table has vertical guides attached to the table and horizontal guide lines machined in 6 supports for the positioning of the first spacers. Plates and spacers are piled up and screwed together. Figure 8 shows a detail of one corner of the assembly of the first plates.

Figure 8: Table for the mechanical structure assembly. It contains horizontal and vertical guide lines.

Figure 9: Detail of one corner during the assembly of the first planes.

Figure 10 shows a picture of the mechanical structure almost finished. Once the structure was completed it was placed (Figure 11) in a rotation tool specially designed and it was rotated (Figure 12) to vertical position. This rotation tool serves not only to rotate the mechanical structure but also the full prototype once it is equipped with the detector electronics and cables. This could be useful to change the orientation to test it in beam tests (vertical) and cosmic rays (horizontal).

The structure deformation has been checked during the assembly and after rotation using a laser interferometer and a 3D articulated arm.

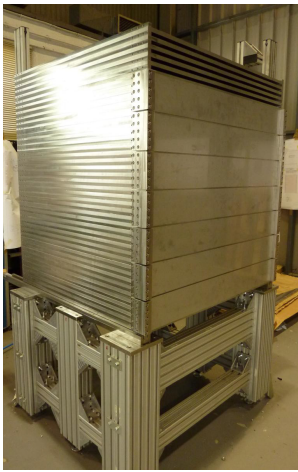
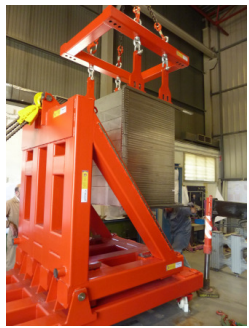
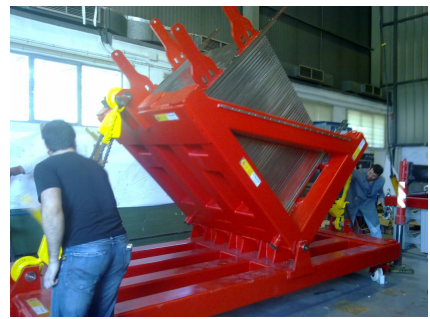

Figure 10: Mechanical structure almost finished.

Figure 11: Mechanical structure being placed on the rotation tool.

Figure 12: Mechanical structure during the rotation from horizontal to vertical position.

LCWS11

### 5.3 The Detector mechanical structure: The cassette

The GRPC detector together with its associated electronics is hosted into a special cassette which protects the chamber, ensures a good contact of the readout board with the anode and simplifies the handling of the detector. The cassette (see Figure 13) is a box made of 2 stainless steel plates 2.5 mm thick and 6 x 6 mm$^2$ stainless steel spacers machined with high precision closing the structure. One of the two plates is 20 cm larger than the other. It allows fixing the three DIFs and the detector cables and connectors (HV, LV, signal cables). A polycarbonate spacer, cut with a water jet, is used as support of the electronics; it fills the gap between the HARDROC chips improving the rigidity of the detector. A Mylar foil (175 mm thick) isolates the detector from the box. The total width is 11mm, 6 of them correspond to the GRPC and electronics, and the rest is absorber.

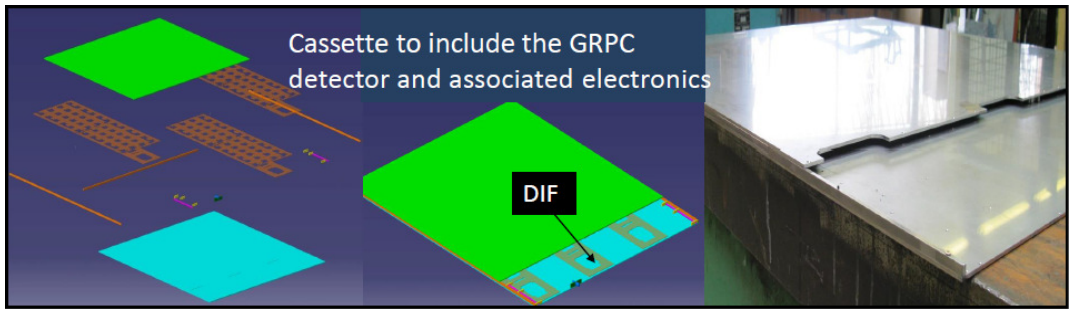

Figure 13: Exploded view of the different cassette components (left). Cassette external view drawing (center). Picture of an empty GRPC cassette (right).

### 5.4 Integration of the GRPC in the structure

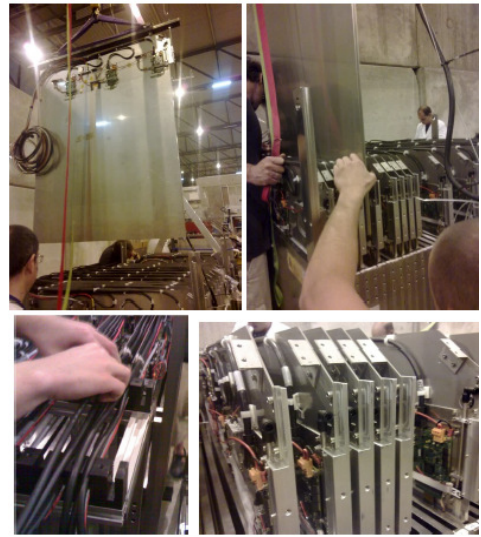

A total of 48 GRPC cassette detectors have been built and inserted in the mechanical structure. The insertion has been made from the top with the help of a small crane as it is illustrated in Figure 14. Vertical insertion minimizes the deformation of the cassettes making easier the procedure. Each cassette is connected to a 5 cables. Three of them correspond to the HDMI readout connections, and the others carry the high and low voltage. The gas is distributed individually to each GRPC. Figure 15 shows a detail of the external cabling distribution in the cassettes.
 After the final assembly several cassettes have been extracted and replaced by others, the operation was done smoothly without problems.

Figure 14: Several pictures showing the insertion of GRPC inside the mechanic structure and final cabling.

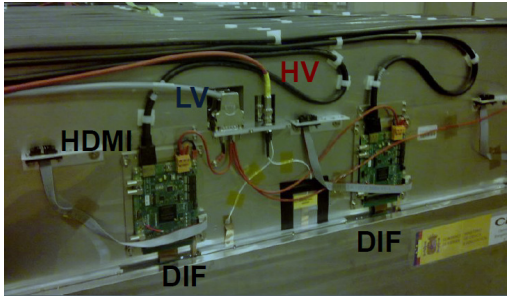

Figure 15: Detail of the cassette cabling. The third DIF is not seen in this picture.

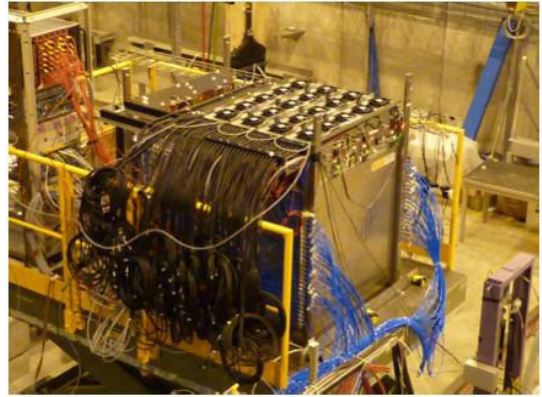

Figure 16: Final prototype at the SPS test beam area.

## 5.5 Prototype Commissioning

The prototype has been exposed to muon and pions beams at CERN. Figure 16 shows a picture of the final prototype at SPS. It was the first attempt to show that both, detectors and electronics perform well, but the new CALICE DAQ generation that is needed to operate this prototype was not completely ready to work properly with this large number of channels. Most of the data taking period was invested on debugging and improving the DAQ system, and more work is ongoing now with cosmic rays. Nevertheless the first results look very promising. Figure 17 shows a typical event display of a muon crossing the SDHCAL prototype. Figure 18 shows the shower development in the prototype for a single pion event (left) and two pions (right) crossing the calorimeter at the same time. Each color corresponds to a different threshold and corresponds to raw data. The results show the details of the shower with a low noise level.

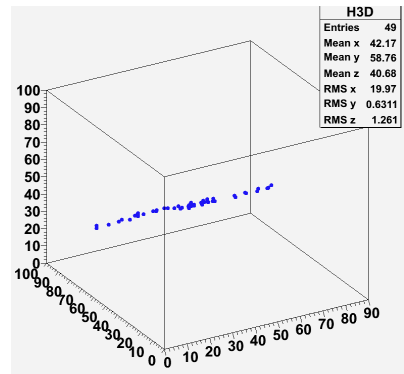

Figure 17: 3D muon event display.

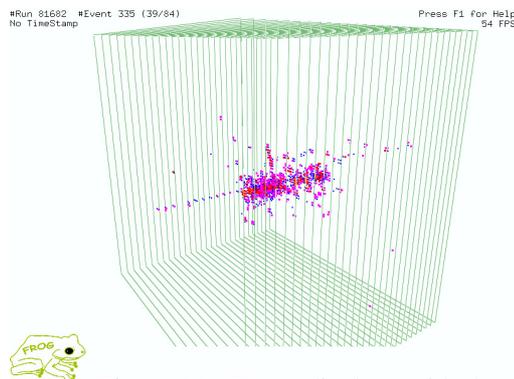
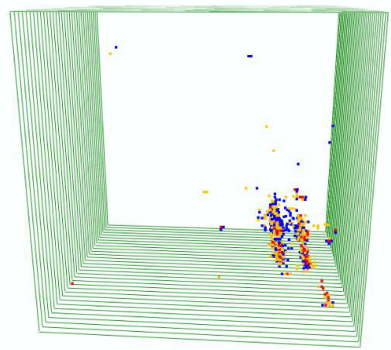

Figure 18: Event displays with the shower development for one (left) and two (right) pions. Different colors corresponds to different thresholds.

## 6 Summary

A self supporting 1m$^3$ SDHCAL prototype using GRPC as active medium has been designed and built. It is made by a mechanical structure containing 51 stainless steel absorber plates and, at present, it is instrumented with 47 detectors. Those detectors are equipped with read out electronics embedded in the detectors and including a power pulsing system. The assembly methods have been tested and no major problems have been found. The very preliminary results are rather promising. The prototype should be exposed to new test beam campaigns during 2012 in order to study the performance of the prototype with enough detail to conclude whether or not it might be considered a valid technology for the hadronic calorimeter of a future linear collider.